\documentclass[a4paper,12pt,useAMS,usenatbib]{mn2e}

\usepackage{color}
\usepackage{graphicx}
\usepackage{aas_macros}
\usepackage{times}
\usepackage{amsmath,amsfonts,amssymb}
\usepackage[
    a4paper,
    plainpages=false,
    breaklinks=true,
    bookmarks=true,
    bookmarksopen=true,
    bookmarksopenlevel=1
]{hyperref}

\def\lesssim{\mathrel{\hbox{\rlap{\hbox{\lower4pt\hbox{$\sim$}}}\hbox{$<$}}}}
\def\gtrsim{\mathrel{\hbox{\rlap{\hbox{\lower4pt\hbox{$\sim$}}}\hbox{$>$}}}}

\def\gsim{\mathrel{\hbox{\rlap{\hbox{\lower4pt\hbox{$\sim$}}}\hbox{$>$}}}}

\title[Heating of the IGM by HMXBs]
      {Heating and Ionization of the Primordial Intergalactic Medium by High Mass X-ray Binaries}
        \author[G. Knevitt, G. Wynn, C. Power  \& J.S. Bolton]
	     { G. Knevitt$^{1}$ \thanks{gillian.knevitt@cantab.net},
               G. A. Wynn$^{1}$, C. Power$^{2,3}$ \& J. S. Bolton$^{3,4}$\\
\\
$^1$Department of Physics and Astronomy, University of 
                Leicester, Leicester LE1 7RH, United Kingdom\\
               $^2$ International Centre for Radio Astronomy Research,
                University of Western Australia, 35 Stirling Highway, Crawley, 
                Western Australia 6009, Australia\\
               $^3$ ARC Centre of Excellence for All-Sky Astrophysics (CAASTRO), 
                University of Sydney, NSW2006, Australia\\
                            $^4$ School of Physics and Astronomy, University of Nottingham, University Park, Nottingham, NG7 2RD, United Kingdom}

\begin{document}

\date{} 

\pagerange{\pageref{firstpage}--\pageref{lastpage}} \pubyear{2012}

\maketitle

\label{firstpage}

\begin{abstract}

We investigate the influence of High Mass X-ray Binaries on their high redshift environments. Using a one-dimensional radiative transfer code, we predict the ionization and temperature profiles surrounding a coeval stellar population, composed of main sequence stars and HMXBs, at various times after its formation. We consider both uniform density surroundings, and a cluster embedded in a  $10^8 M_{\odot}$ NFW halo. HMXBs in a constant density environment produce negligible enhanced ionization because of their high-energy SEDs and short lifetimes. In this case, HMXBs only marginally contribute to the local heating rate. For NFW profiles, radiation from main sequence stars cannot prevent the initially ionized volume from recombining since it is unable to penetrate the high density galactic core. However, HMXB photons stall recombinations behind the front, keeping it partially ionized for longer. The increased electron density in these partially ionized regions promotes further cooling, resulting in lower IGM temperatures. In the context of this starburst model, we have shown that HMXBs do not make a major contribution to reionization or IGM heating. However, X-ray escape fractions are high in both density profile cases. Continuous star formation may result in the build up of X-rays over time, reducing the ionization timescale and potentially leading to low level ionization of the distant IGM.
\end{abstract}



\section{Introduction}
\label{sec:intro}


X-ray radiation may have significantly influenced its surroundings in the distant past, shaping the universe we see today. For example, the presence of an X-ray background at high redshifts could have increased the fractional ionization of protogalactic gas, offsetting the effects of photodissociation from UV feedback, and allowing gas haloes to cool at higher virial temperatures \citep{2000ApJ...534...11H, 2001ApJ...553..499O, 2003MNRAS.338..273M}. In addition, the cosmological reionization of neutral hydrogen, which completed approximately one billion years after the Big Bang \citep[see reviews by e.g.][]{2001PhR...349..125B, 2010Natur.468...49R}, may have been profoundly influenced by the presence of X-rays. Although UV radiation from massive stars is likely to have been the main contributor to reionization \citep{2004MNRAS.350...47S,2004ApJ...605..579Y, 2008ApJ...684....1W}, X-rays may have been able to ``pre-ionize" large volumes of the IGM, unreachable by UV sources \citep{2001ApJ...553..499O, 2001ApJ...563....1V, 2004ApJ...604..484M, 2004MNRAS.352..547R}. Their penetrative power has lead some authors \citep[e.g. ][]{2003MNRAS.340..210G, 2004ApJ...604..484M,2011Natur.472...47H} to propose a smoother global increase in ionized fraction with a more uniform morphology than the currently favoured "Swiss Cheese Paradigm" \citep{2000ApJ...535..530G,2001PhR...349..125B,2007MNRAS.377.1043M}, whereby ionized regions formed in isolated pockets that eventually expanded and joined. Heating from X-rays could also affect the progress of reionization, delaying its onset by increasing the Jeans mass \citep{2004MNRAS.352..547R,2005MNRAS.363.1069K}. \citet{ 2013MNRAS.431..621M} show that the presence of X-rays may result in an extended epoch during which hydrogen is $\sim$10$\%$ ionized, before UV photons complete the process. Finally, X-ray radiation could further influence reionization by suppressing small scale structure formation, thereby lowering recombinations \citep{2013arXiv1310.7944J}.

However, the exact effects of X-ray feedback in the early universe are unclear, and depend on both the level of radiation produced, and the sources responsible for it. Several observations can be used to constrain early X-ray emission. For example, the high-redshift abundance of hard ($>$$1$keV) X-ray photons is limited by the unresolved soft X-ray background (SXB), since they were optically thin to the IGM and remain unabsorbed to the present day \citep{2004ApJ...613..646D, 2007MNRAS.374..761S,2012MNRAS.426.1349M}. These high energy X-rays would have also altered the ionization state of metal absorption lines in $z\sim 3$ quasar spectra \citep{2012MNRAS.426.1349M}. Upper limits on the kinetic Sunyaev-Zeldovich signal support homogenous ionization  for which X-rays are an obvious candidate \citep{2012MNRAS.422.1403M,2012ApJ...755...70R,2012JCAP...05..007V}.  Furthermore, the level of X-rays can determine whether the redshifted 21cm spin-flip line appears in absorption or emission \citep{2006PhR...433..181F,2013MNRAS.431..621M}. Feedback from X-rays must also conform to measurements of the thermal state of the IGM \citep{2011arXiv1110.0539B,2012MNRAS.421.1969R}, and timescales for He$^+$ ionization \citep{2003ApJ...596....9H,2009ApJ...706..970D,2009ApJ...694..842M, 2011MNRAS.410.1096B,2011ApJ...733L..24W}. 

Feedback from several X-ray emission mechanisms have been evaluated within these constraints, including accretion onto super-massive black holes \citep[e.g.][]{2004MNRAS.352..547R, 2005MNRAS.357..207R, 2009ApJ...703.2113V}, shocks in supernova remnants \citep{2011ApJ...743..126J},  dark matter annihilations \citep{2009PhRvD..80c5007B} and accretion of the ISM onto isolated stellar mass black holes \citep{2011ApJ...738..163W}.

A source of high-redshift X-ray radiation that has gained particular attention in recent years is that of stellar-mass X-ray binaries. These sources are now believed to be proportionally more luminous at higher redshifts than the present day, due to an anti-correlation of HMXB mass with metallicity \citep{2006MNRAS.370.2079D, 2010MNRAS.403L..41C,2011A&A...528A.149M}. In fact, \citet{2013ApJ...764...41F} have argued that these systems are likely to dominate X-ray luminosities over AGN at $z\gsim$ 6 - 8. 

Several previous studies have highlighted the potential influence of X-ray binaries on their surroundings.  For example, their feedback effects have been addressed by  \citet{2003MNRAS.340..210G} and \citet{2012MNRAS.423.1641J}, the latter of whom predict a significant, stochastic effect on the evolution of dwarf galaxies, which could be responsible for the diversity we see today. High Mass X-ray Binaries (HMXBs) have been evaluated as a source of reionization in our previous work \citet{2009MNRAS.395.1146P}, and by \citet{2011A&A...528A.149M}. Both studies suggest that the increased mean free path of X-rays and their potential for secondary ionizations lead to the ionizing potential of HMXBs equalling that of their progenitors. The X-ray energy emitted by HMXBs in the early universe, relative to stellar radiation alone, has been quantified in both \citet{2013ApJ...776L..31F} and our most recent study, \citet{2013ApJ...764...76P}. 

Qualitative reasoning and calculations of the energetics imply a significant (up to $10\%$) contribution from HMXBs to reionization. However, \citet{2012MNRAS.426.1349M} argues, based on assuming a hard power-law spectrum for HMXBs, that they cannot have contributed to reionization without the SXB constraint being violated. \citet{2014MNRAS.tmpL..23K}, on the other hand, predicts a much weaker constraint from the SXB, because HMXB show spectral curvature above 2keV. In another study, \citet{2013arXiv1310.7944J} computed the effects of the very first HMXBs on their environments using a zoomed hydrodynamical simulation, incorporating a fully integrated radiative transfer model. They found little or no direct ionization of the IGM from X-rays, although they predicted a net positive effect on reionization due to the suppression of small scale structure. The discrepancy between these works, and among general HMXB ionization predictions, highlight the need to move beyond energetics arguments when assessing the feedback potential of HMXBs; full radiative transfer calculations, with accurate HMXB spectra, are required to predict the proportion of energy that directly contributes to heating and reionization, and the timescales over which it acts. 

Our Monte Carlo stellar population models, with observationally motivated HMXB abundances, lifetimes, and, importantly, spectra, provide the ideal basis for an improved study of the influence of HMXBs within a population of stars. We have previously shown \citep{2013ApJ...764...76P} that our models do not conflict with the SXB constraint. We now use one dimensional radiative transfer calculations to estimate the influence of HMXBs in a realistic stellar cluster on their environment, with no presuppositions about their significance to reionization.

In Section 2 we describe our stellar population simulations and our radiative transfer calculation. We then use these models in Section 3 to predict the ionization and temperature profiles surrounding a high-redshift stellar cluster, with and without a HMXB component, in both constant and varying density environments.  In Section 4, we discuss the factors that influence our results, and the conditions required for HMXBs to have a strong influence on their surroundings, before drawing our final conclusions in Section 5. 

\section{Methods}

We use the starburst model we developed in \citet{2009MNRAS.395.1146P,2013ApJ...764...76P}, to simulate a stellar population forming in an instantaneous burst, containing both main sequence stars and HMXBs. We monitor the total energy released from this cluster and enter it into a one-dimensional time-dependent radiative transfer code, first described in \citet{2007MNRAS.374..493B}, to estimate the ionization state and temperature of the surroundings. These models are described below. 

\subsection{Stellar Population Synthesis}

\begin{figure}
\centering
\includegraphics[angle = -90, width= 0.6\textwidth]{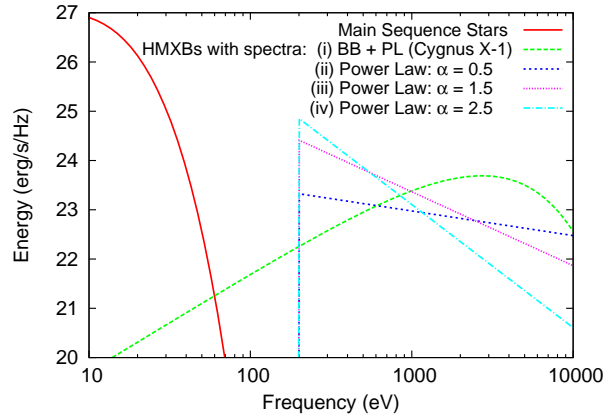}
\caption{\textbf{Choice of HMXB Spectrum.} Example spectrum snapshots from our Monte Carlo stellar population models, taken at 20 Myr, assuming a Kroupa IMF and $f_{\rm sur} = 1$. The combined spectrum from all the main sequence stars is plotted in red. The other lines show the net spectral output from the HMXBs for four different choices of HMXB spectrum. Our fiducial black body and hard power law spectrum, based on that of Cygnus X-1, is plotted as a light green dashed line. Power law HMXB spectra with $\alpha$ = 0.5, 1.5, and 2.5 are plotted as dark blue dotted, purple dotted and light blue dot-dashed lines respectively.}
\label{fig:spectra}
\end{figure}

A full description of our model can be found in \citet{2009MNRAS.395.1146P}. We briefly describe its main features, including our parameter choices, below.

Initially $10^6$ stars are formed instantaneously, following a \citet{2001MNRAS.322..231K} mass distribution, with stellar masses in the range $0.01 \leq {\rm M_{\ast}/M_{\odot}} \leq 100$. We assume all stars are formed in pairs, and binary parameters are then assigned following the approach of \citet{2006MNRAS.370.2079D}, with orbital periods distributed uniformly in logarithm between 1 and $10^4$ days. Stellar lifetimes are estimated using \citet{2001A&A...371..152M,2003A&A...399..617M}, with the assumption of Z = 0.

Stellar luminosity $L_*$ follows a mass-luminosity relation of the form,
\begin{equation}
  \label{eq:mass-lum}
  \frac{L_{\ast}}{L_{\odot}}=\alpha \,\left(\frac{\rm M_{\ast}}{\rm M_{\odot}}\right)^\beta,
\end{equation}
\noindent where the constants $\alpha$ and $\beta$ depend on stellar mass and can be found in Table 1 of \citet{2013ApJ...764...76P}. Stars then radiate as blackbodies with \begin{equation}
  \label{eq:teff}
  T_{\rm eff}=\left(\frac{L_*}{4\pi\,R_{\ast}^2\sigma_{\rm SB}}\right)^{1/4}
\end{equation}
\noindent where the stellar radius, $R_{\ast}/R_{\odot}=(\rm M_{\ast}/\rm M_{\odot})^{0.8}$ and $\sigma_{\rm SB}$ is the Stefan-Boltzmann constant. The cumulative SED from all of the main sequence stars in our model is plotted 20 Myr after the birth of the cluster as a solid red line in Figure \ref{fig:spectra}. 

HMXBs form when a primary ends its main sequence life, provided several conditions are met. Firstly, the initial mass of the primary must exceed $8 \rm M_{\odot}$, the threshold for neutron star formation \citep[cf. Figure 1 of][]{2003ApJ...591..288H}. Next the secondary must have $M_* \gsim 3 M_{\odot}$; this fits the HMXB and intermediate mass binary definitions. When the primary of a system fitting these conditions leaves the main sequence branch, we estimate the remnant mass using Figure 3 of \citet{2003ApJ...591..288H}, and calculate the revised binary parameters. If the system loses more than half of its mass in its supernova phase, it becomes unbound so is no longer a HMXB candidate; this removes $70\%$ of all systems. A further $f_{\rm sur}$ of the population then survives, where, in this study we assume $f_{\rm sur} = 1$. This parameter is discussed in detail in \citet{2009MNRAS.395.1146P}

We draw HMXB luminosities from a Weibull distribution, with a peak luminosity of $L_{\rm X} \sim 10^{38}$ 
erg/s, capped at an upper limit of $L_{\rm X} \simeq 1.26 \times 10^{38} (\rm M/\rm M_{\odot})$ erg/s so that they do not accrete at super-Eddington rates. We then consider four alternative HMXB spectra:

\begin{enumerate}
\item For our fiducial model, we base HMXB spectra on the galactic HMXB, Cygnus X-1, which alternates between a high luminosity \textit{soft state} and and low luminosity \textit{hard state}. In our model, if $L_X \gsim 10^{37}$ erg s$^{-1}$, we assume a soft state, composite black body and power-law spectrum \citep[cf. ][]{1977Natur.267..813D, 1982Natur.295..675O}.  The blackbody temperature is 
  calculated by assuming ($L_{X}/L{_{\rm Cyg X-1}}$) = ($T/T{_{\rm Cyg X-1}}$)$^4$, and the power-law has a slope of $\alpha =  -1.1$, normalised such that the ratio between the blackbody and power law components matches that of Cygnus X-1. For $L_X < 10^{37}$ sources, we model a hard state power-law spectrum with $\alpha = -0.8$ between 2 and 10keV \citep[cf. ][]{1997MNRAS.288..958G,1999MNRAS.309..496G}. 

The total SED of the HMXBs in the cluster at 20 Myr is plotted as a green dashed line in Figure \ref{fig:spectra}. The contribution from hard state spectra is negligible. As such, this SED represents a sum of the most luminous soft state HMXB composite black body and power law spectra. This model will be subsequently be referred to as a BB+PL spectrum.  
 \item A power law spectrum,  $F(E) = E^{-\alpha}$, and $\alpha = 0.5$ in the observed X-ray range of 0.2-30keV, where the upper cut-off energy is based on INTEGRAL observations of HMXBs from  e.g.  \citet{2005A&A...444..821L}. The cumulative SED of all HMXBs in this model after 20 Myr is plotted in Figure \ref{fig:spectra} as a dark blue dotted line. 
\item A power law spectrum; $\alpha = 1.5$, plotted in Figure \ref{fig:spectra} as a  fine purple dotted line.
\item A power law spectrum; $\alpha = 2.5$,plotted in Figure \ref{fig:spectra} as a light blue dot-dashed line.
\end{enumerate}

\noindent Our model output is a global time-dependant SED, which is then used as an input for our radiative transfer calculations.

\subsection{1d Radiative Transfer Code} \label{sec:1dRT}

Our line-of-sight radiative transfer implementation is based on the algorithm described in detail in Appendix B of \citet{2007MNRAS.374..493B}, to which we refer the interested reader for further details.  It is, itself, based on the radiative transfer algorithm originally developed by \citet{1999ApJ...523...66A}, which was subsequently extended to incorporate a multi-frequency treatment by \citet{2004MNRAS.348L..43B}.   Here we limit ourselves to briefly describing the updates to the code relevant for this work.

The algorithm follows the radiative transfer of ionising photons with energies in the range $13.6\rm eV < E < 4\rm keV$ through a medium consisting of hydrogen and helium, with the option of using up to 80 separate photon energy bins over this range.  Secondary ionisations by fast electrons are included, using the recent results presented by \citet{2010MNRAS.404.1869F}. The absorption probabilities for ionising photons have furthermore been updated to match the form recently recommended by \citet{2012MNRAS.421.2232F}. Finally, we have tested this code against the analytical result for an expanding H$^+$ region in a uniform medium \citep[see also][]{2007MNRAS.374..493B}, as well as the recent results of \citet{2012MNRAS.426.1349M} (their Fig 5), confirming the code yields consistent results in both cases.

\section{Results}
\begin{figure*}
\centering
\includegraphics[angle = -90, width= 1.03\textwidth]{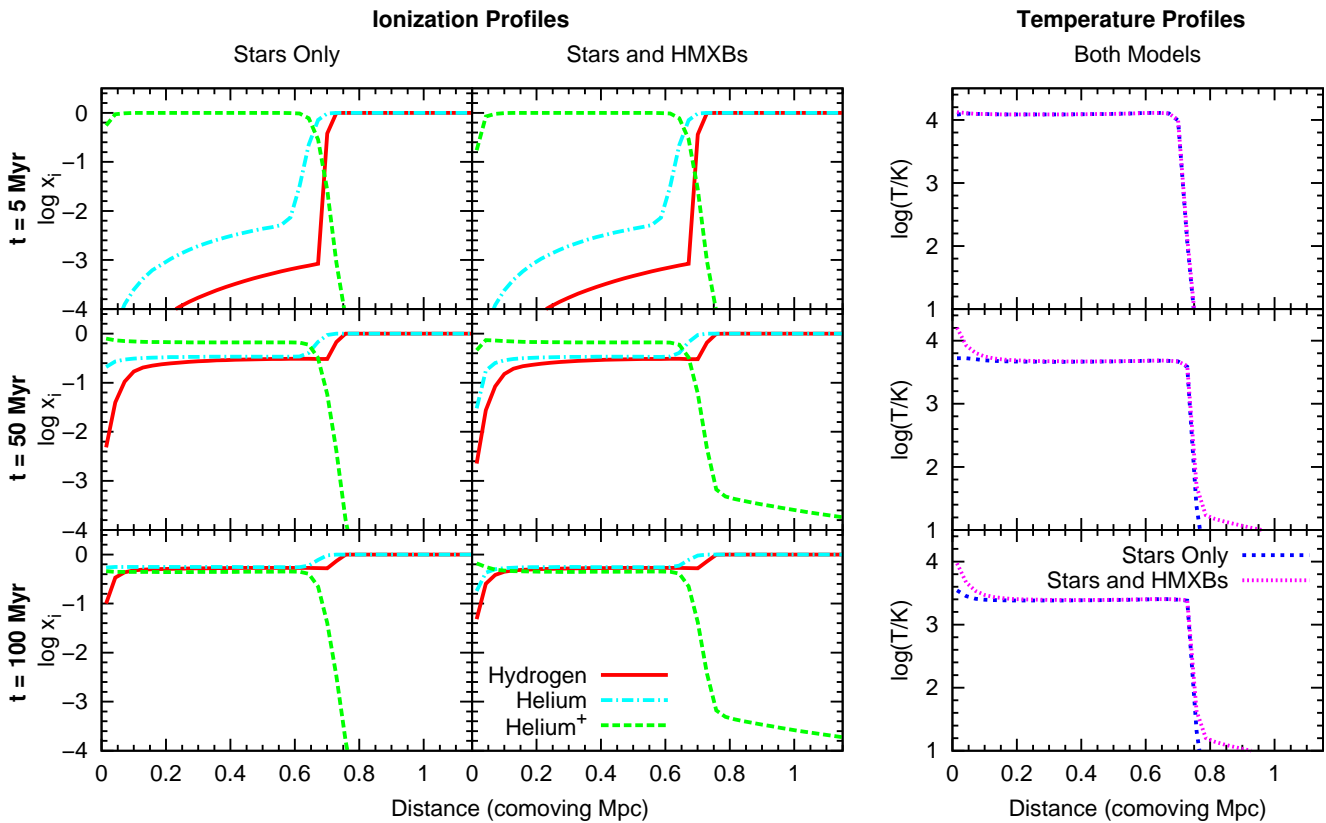}
\caption{\textbf{Heating and Ionization from a model stellar cluster surrounded by a uniform density medium: including a HMXB phase.} These plots show the ionization and temperature profiles surrounding our simulated stellar populations with and without HMXBs. Profiles were calculated using our one dimensional time dependent radiative transfer code, assuming a uniform density medium equal to the redshift dependent average background density, for a 200 Myr simulation ending at z = 10.  In the left-hand and middle columns, the hydrogen neutral fraction (solid red line), helium neutral fraction  (light blue dot-dashed line) and helium$^+$ abundance (green dashed line) are plotted against distance. These are shown at 5, 50 and 100 Myr after the birth of the stellar cluster in the upper, middle and lower panels respectively. The output from a model with only main sequence stars ($f_{\rm sur}= 0$) is plotted on the left while the panels in the middle column show results from a model comprising both stars and HMXBs ($f_{\rm sur} = 1$). The right-hand panels plot temperature against distance for both models; the stellar-only simulation is represented by a dark blue dotted line and the model containing a further HMXB phase is plotted as a fine purple dotted line. }
\label{fig:const_density_ion_temp}
\end{figure*}

The ionizing energy output of a single stellar population is initially dominated by its most massive stars, until they end their short main sequence lives. However, this may not be their only contribution to the ionizing power of the cluster;  if they evolve to become accreting black hole systems, they can then provide further ionizing energy to their surroundings. Figure 2 of \citet{2013ApJ...764...76P} shows the stellar cluster's total ionizing energy ($13.6$eV $<$ E $<$ $1$keV) against time, separated into a stellar component (red line) and a HMXB component (all other lines), where the energy produced by the HMXB component depends on the fraction of main sequence binaries that evolve to become HMXBs ($30\% {\times}$ $f_{\rm sur}$; cf. Section 2.1).  In this study we showed that the ionizing energy from HMXBs becomes dominant after $\sim$ 20 Myr, provided $f_{\rm sur} \gsim 0.5$.

In the following section, we use the time-dependent SED of the same model cluster to study the ionization and heating of its surroundings. We compare our results for models with no post-MS systems to those with HMXBs ($f_{\rm sur} = 1$), and comment on the additional effect a HMXB phase may have on the cluster's environment.

\subsection{Heating and Ionization of  a Constant Density Medium} \label{sec:const_dens}

In the first case, we consider the propagation of ionizing energy from our model cluster directly into the IGM, using our one dimensional radiative transfer code described in Section \ref{sec:1dRT}. We assume a uniform IGM density equal to the average matter density at a given redshift. We then follow the cluster's emission from its birth, at z = 14.5, for 200 Myr, until  z = 10. The IGM density decreases over this time due to cosmological expansion.

 I estimate $E_{\rm lim}$ in the same way as for \citet{2009MNRAS.395.1146P}, by requiring that $\sigma(E_{\rm lim})=(H(z)/c)/n(z)$ where $\sigma(E)$ is the ionization cross section of neutral hydrogen, $H(z)$ is the Hubble parameter at redshift $z$, $n(z)$ is the mean baryon density and $c$ is the speed of light. For $z \gtrsim 10$, $E_{\rm lim} \gtrsim 1$ keV. 
\subsubsection{Ionization Profiles}

Snapshots from our radiative transfer calculation, at 5, 50 and 100 Myr (top, middle and bottom rows) after the birth of the cluster, are plotted in Figure \ref{fig:const_density_ion_temp}. In the left-hand and middle columns, abundances are plotted against distance for a cluster with no HMXBs (left) and one with a HMXB phase (middle). The neutral hydrogen fraction is plotted as a function of distance from the cluster, as a solid red line. Similarly the neutral helium fraction and He$^+$ abundance are plotted as blue dot-dashed and green dashed lines respectively.

In the stellar-only models (left column) the hydrogen and helium ionization fronts begin by expanding rapidly into the gas. As the number of ionizing photons close to the outer edges drops, recombinations begin to stall the ionization fronts, so that their maximum extent, at 5 Myr, is $\sim 0.7$ Mpc. Since the recombination rate is fairly uniform within the fully ionized zone, the neutral fraction then gradually increases to $\sim$30\% at 50 Myr.  By 100 Myr, the neutral fraction of the gas behind the ionization fronts has increased to $\sim 50\%$. An inner highly ionized region close to the source of radiation exists for hydrogen at both 50 and 100 Myr, though at later times it has significantly reduced in extent. A similar feature can be seen in the helium profile, which disappears by 100 Myr. After 100 Myr, $\sim$40$\%$ of the helium within 0.7 Mpc is He$^{+}$,  $< 0.1\%$ is He$^{2+}$, and the remainder is neutral. 

The inclusion of HMXBs (middle column) has very little effect on any of the ionization profiles. Close to the source of radiation ($<$ 100 kpc), there is some additional ionization of hydrogen and helium at late times, but the extent of the main ionization fronts are identical to the purely stellar case. At early times, the harder photons from X-rays cause a very local ($<20$kpc) ionized He$^{2+}$ region close the cluster, which recombines at later times.

\subsubsection{Temperature Profiles}

Also plotted in Figure \ref{fig:const_density_ion_temp}, in the right-hand panels, are the temperature profiles surrounding the two model clusters at three different times; 5, 50 and 100 Myr (top to bottom panels) after the birth of the cluster. The profile for a main sequence stellar-only model is shown as a dark blue dashed line, and the model with a HMXB phase is plotted with a fine dotted purple line. Within 5 Myr, the region behind the ionization front is heated to $\sim 10^4 \rm K$, and cools slowly, reaching $\sim$ 2500K after $100$ Myr. The temperature profiles with and without HMXBs are similar in the main ionized zone, but after 50 Myr, they diverge in both the innermost region and just outside the ionization front. HMXBs are responsible for increased heating local to the source of radiation due to hydrogen and helium photo-heating close to the source. At large distances, beyond the ionization front, a slightly extended heated region is visible, associated with the low level of He$^+$ photo-heating by hard X-ray photons at these distances. 

\subsubsection{Different Spectra}

\begin{figure}
\centering
\includegraphics[width=0.5\textwidth]{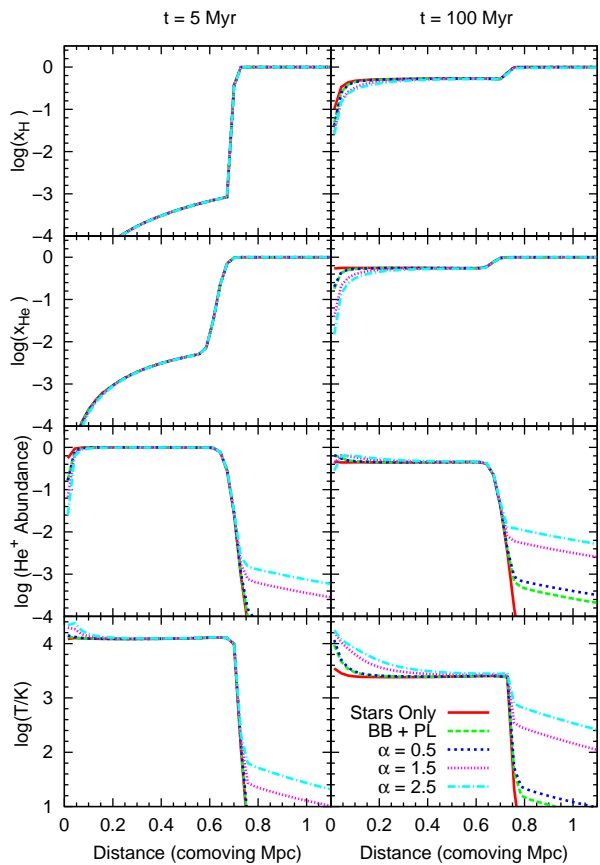}
\caption{\textbf{Heating and Ionization from a model stellar cluster surrounded by a uniform density medium: choice of HMXB spectrum.} Plots comparing the effect of our choice of HMXB spectrum on the ionization and temperature profiles surrounding a model stellar cluster embedded in a uniform density medium. Calculations were made using our one dimensional radiative transfer code, inputting the radiation from the cluster over 200 Myr, ending at z = 10. The left and right-hand panels show results at 5 and 100 Myr respectively after the birth of the cluster. Plotted as a solid red line is the output for a stellar-only simulation. Models including HMXBs are then plotted  for a black body + power law spectrum (dashed green line) and three power law spectra; $\alpha = 0.5$ (dark blue dotted line), $\alpha = 1.5$ (fine purple dotted line) and $\alpha = 2.5$ (light blue dot-dashed line). From top to bottom, the panels show the hydrogen neutral fraction, helium neutral fraction, helium$^+$ abundance and gas temperature as a function of distance from the cluster. }
\label{fig:const_density_compare_spectra}
\end{figure}

The ionizing and heating potential of HMXBs will depend on their spectra. HMXBs with our fiducial model spectrum, a composite blackbody and high energy power law (BB+PL), had negligible influence on their surroundings when added to the radiation already present from main-sequence stars. We now test the alternative, power-law, spectra shown in Figure \ref{fig:spectra}, with $\alpha = 2.5, 1.5$ and $0.5$. Since the power law spectra extend from 200eV to 30keV, different slopes lead to different proportions of HMXB energy falling within the useful ionizing range, from E$_{\rm min}$ = $13.6\rm eV$, to E$_{\rm max} \approx$ a few keV, above which, photon mean free paths are too long for significant interaction with the local IGM. In fact, while 98$\%$ of the energy for $\alpha = 2.5$ falls within the 13.6-4keV energy range used in our radiative transfer code, just $84.5\%$ and $30.9\%$ of the HMXB energy from spectra with $\alpha$ = 1.5 and 0.5 respectively fall within these bounds.

Figure \ref{fig:const_density_compare_spectra} shows the resulting ionization and temperature profiles generated by re-running our one dimensional radiative transfer calculation with these different HMXB spectra. Results are shown at 5 Myr and 100 Myr on the left and right panels respectively. On the upper panels, the hydrogen neutral fraction is plotted as a function of distance for five different models; our two previous models, without (solid red line) and with (green dashed line) HMXBs, assuming a BB+PL spectrum, and our three new power-law models with spectral slopes, $\alpha$, of 0.5 (dark blue dotted line), 1.5 (fine purple dotted line) and 2.5 (light blue dot-dashed line). At 5 Myr, there is no discernible difference in the ionization state of hydrogen between the models, as stellar radiation still dominates the ionizing power. At 100 Myr, there are minor differences depending on HMXB spectra within 100 kpc of the radiation source. HMXBs with softer spectra are able to keep these nearby regions marginally more ionized than those with harder spectra. On the next row, the same results are shown for helium ionization, where HMXBs have the same effect; larger ionized 
fractions within 100 kpc for softer spectra. He$^+$ abundance is plotted against distance on the third row. At 5 Myr, the local He${^+}$ ionization is higher for softer spectra. At both 5 and 100 Myr, some singly ionized helium is present outside the main ionization front, with low levels of secondary ionization. Here, harder spectra, such the BB+PL model, and a power law with $\alpha = 0.5$, produce higher He$^{2+}$ fractions. 

The temperature profiles for the different HMXB spectra are shown in the lower panels of Figure \ref{fig:const_density_compare_spectra}. Softer power-law spectra, with larger values of $\alpha$, increase the temperature of the inner 0.1 Mpc to a greater extent than the BB+PL model. They also have a visible effect on the temperature outside the ionization front, raising it to $150\rm K$ and $400\rm K$ for $\alpha = 1.5$ and $\alpha = 2.5$ respectively at 1 Mpc from the cluster after 100 Myr. Softer power-law HMXB spectra lead to higher temperatures because a larger proportion of their energy falls below E$_{\rm max}$. HMXBs with particularly soft spectra may therefore be able to extend the volume heated by a typical starburst.

 \subsection{Heating and Ionization from HMXBs within a Galactic Halo}

\begin{figure*}
\centering
\includegraphics[angle = -90, width= 1.03\textwidth]{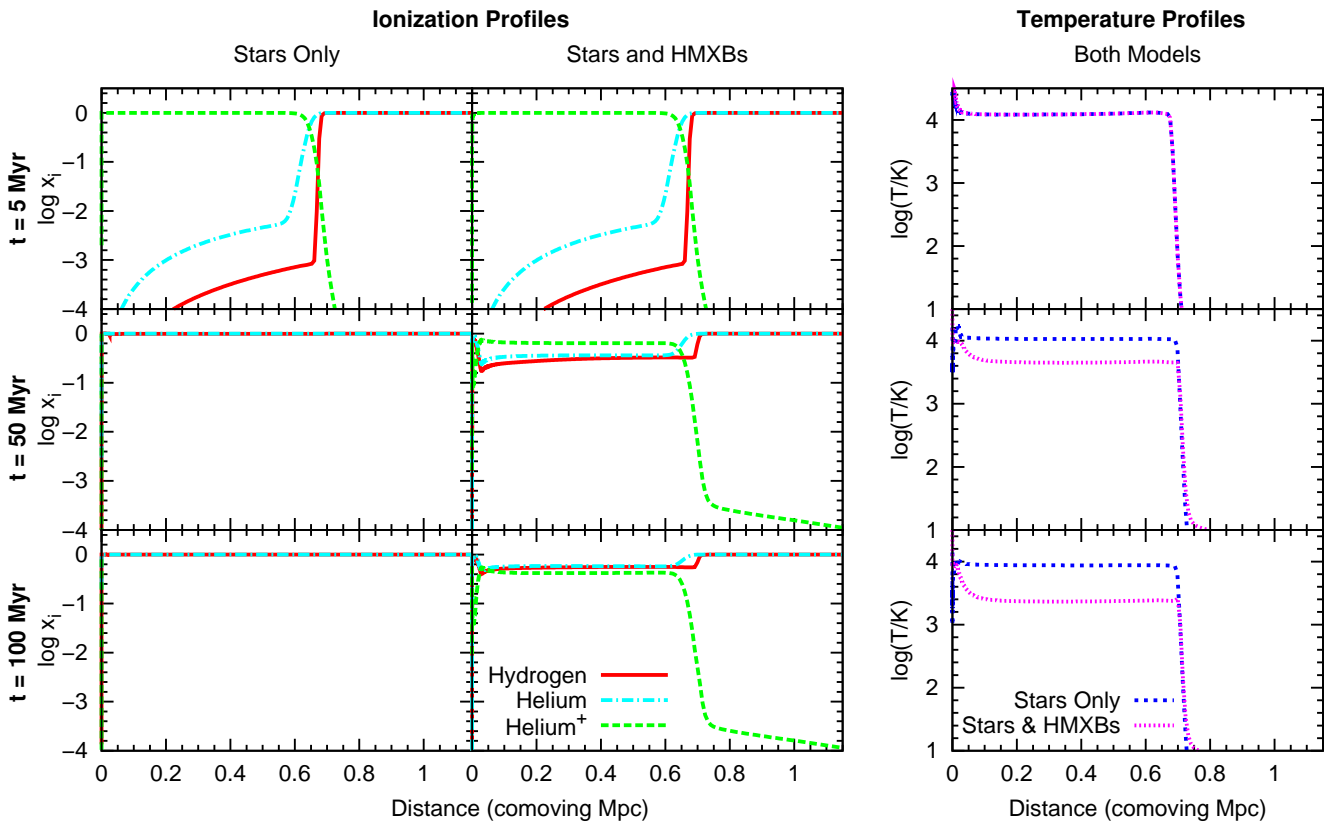}
\caption{\textbf{Heating and Ionization from a model stellar cluster embedded in a $\mathbf{10^8}$ M$_{\boldsymbol{\odot}}$ NFW halo: including a HMXB phase.} Ionization and temperature profiles for a stellar cluster, with and without HMXBs, embedded in a $10^8$ M$_{\odot}$ halo following an NFW density profile. The stellar cluster simulation ran for 200 Myr, ending at z = 10. The radiation from the cluster was then entered into a one dimensional time dependent radiative transfer code to calculate its effect on the surrounding medium. On the left-hand panels, hydrogen neutral fraction (solid red line), helium neutral fraction  (light blue dot-dashed line) and helium$^+$ abundance (green dashed line) are plotted against distance from the cluster for a model containing main sequence stars only. Panels on the upper, middle and lower rows show these profiles at 5, 50 and 100 Myr respectively. In the middle column, the same plots are made for a second simulation that includes HMXBs in addition to stars.  In the right hand panels, the gas temperature has been plotted against distance for the stellar-only simulation (dark blue dotted line) and the model containing HMXBs (fine purple dotted line). }
\label{fig:NFW_density_ion_temp}
\end{figure*}

In a constant density medium, a HMXB phase does not provide any significant additional heating and ionization above that released during the main sequence phase of a single starburst, unless the HMXB spectrum is particularly soft. This result predicts the effect of releasing all of the radiation from a stellar cluster directly into the IGM. Since starbursts form within galaxies, it is unlikely that all of their radiation escapes into the IGM; in a high column density galactic environment, X-rays may be able to penetrate regions that UV radiation cannot. To study the effect of such a density profile, we now use a simple model with a stellar cluster placed at the centre of a galaxy whose gas follows a radially dependent NFW profile out to a fixed radius, beyond which the IGM is at mean density: 

\begin{equation}
\rho( r)= \frac{\rho_{\rm crit} \delta_{\rm char}}{\left(\frac{r}{R_s}\right)^{\alpha}\left(1+\frac{r}{R_s}\right)^{(3-\alpha)}}\label{eqn:nfw}
\end{equation}

Here, $R_s$ is the scale radius; $\rho_{\rm crit}$ is the mean density of the Universe, which we take to be  $2.7755{\times}10^{11}\rm M_{\odot}\rm Mpc^{-3}$; $\delta_{\rm char}$ is the characteristic overdensity; and $\alpha$ is the inner asymptotic slope of the profile.  

For X-rays to propagate further than UV radiation, the central region needs to be dense enough to block the UV, but not so dense that the X-rays are also prevented from escaping. We force this scenario with an example halo with a chosen virial mass of $10^8M_{\odot}$ and a concentration of 10. These parameters define both the virial radius and the scale radius $R_s$. We model a ``cored'' galaxy, with $\alpha = 0$, and assume that the baryon density follows the dark matter density, with a baryon fraction of $16\%$. The halo extends to $\sim 6$ kpc; at greater radii, the medium returns to the constant IGM density used in the last model.

\subsubsection{Ionization Profiles}

In Figure \ref{fig:NFW_density_ion_temp}, we plot the hydrogen, helium and He$^+$ ionization profiles produced by a cluster without and with HMXBs (left and middle panels respectively), assuming a BB+PL HMXB spectrum. These are plotted at 5, 50 and 100 Myr after the birth of the cluster, from top to bottom rows.

In the first 5 Myr, the initial burst of UV radiation from the most massive stars in the cluster dominates its energy output. Since it is able to escape the inner regions of the galaxy, the ionization profiles at this time are the same, with and without HMXBs. In the stellar-only model, the ionized region produced by the most massive main sequence stars has fully recombined by 50 Myr, and the volume remains neutral from then onwards. 

\begin{figure}
\centering
\includegraphics[width=0.5\textwidth]{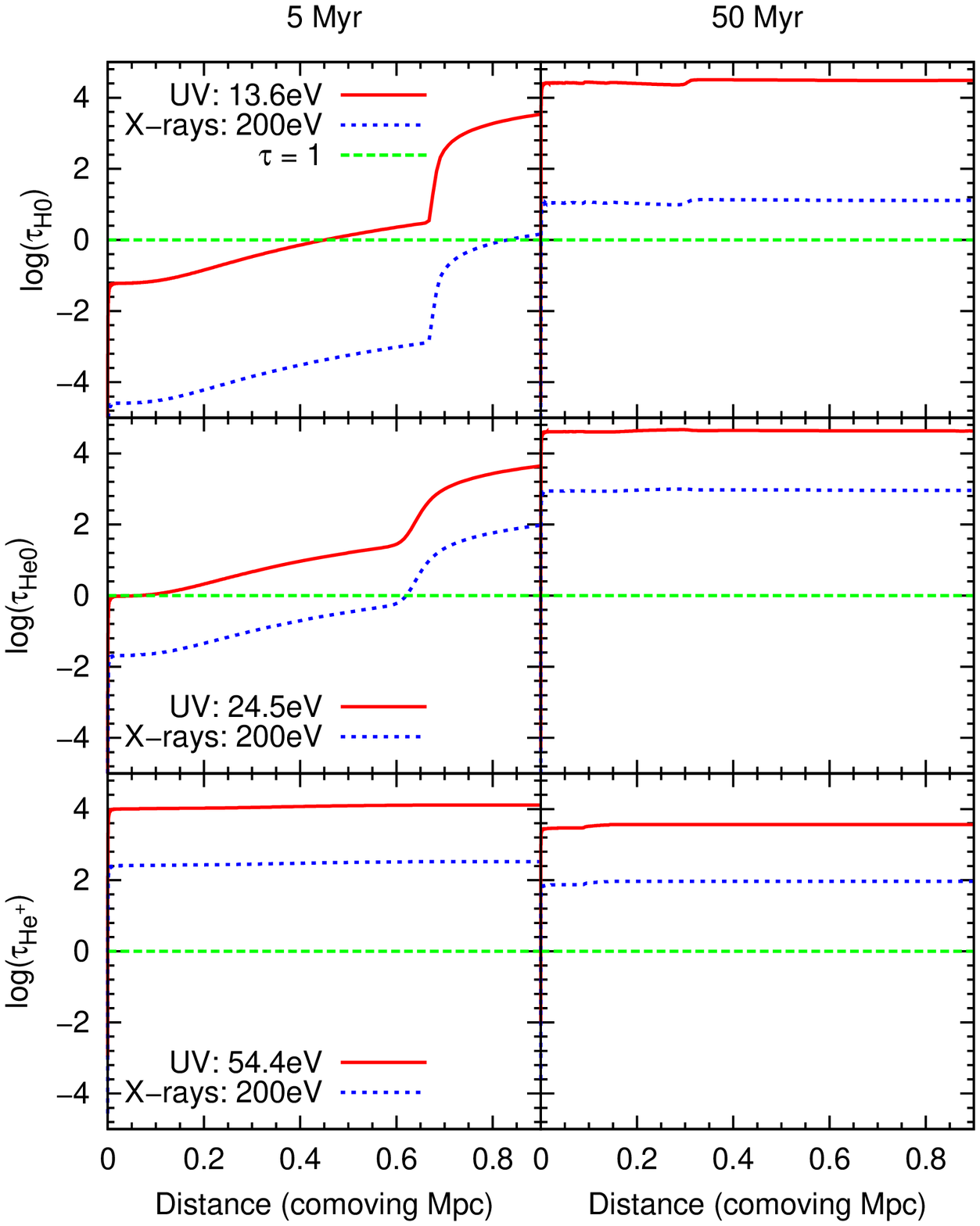}
\caption{\textbf{Heating and Ionization from a model stellar cluster embedded in a $\mathbf{10^8}$ M$_{\boldsymbol{\odot}}$ NFW halo: Optical Depths.} The optical depths of UV (solid red line) and X-ray (dotted blue line) photons are plotted as a function of distance for a stellar cluster containing a HMXB phase embedded in an NFW density profile with M$_{\rm halo} = 10^8$ M$_{\odot}$. The top panels show the optical depth of photons with the UV threshold energy (13.6eV) and X-rays (200eV) to neutral hydrogen at 5 Myr and 50 Myr in the left and right panels respectively. For reference $\tau = 1$ is shown as a green dashed line. Similarly optical depths are shown for neutral helium and He$^+$ in the middle and bottom panels, assuming UV ionization threshold energies of 24.5eV, and 54.4eV respectively, alongside X-ray energies of 200eV. }
\label{fig:tau}
\end{figure}

In contrast, the presence of HMXBs stalls the recombinations, and preserves the ionized region throughout the cluster's lifetime, so that it is still $45\%$ ionized at 100 Myr. This is because the longer mean free path of X-rays from HMXBs allows them to propagate further through their surroundings than UV energy from stars. This is shown in Figure \ref{fig:tau}, where the optical depth of UV and X-ray radiation to hydrogen, neutral helium and He$^+$, are plotted from top to bottom rows as a function of distance. These optical depths are calculated for radiation propagating through the IGM at $5$ and $50$ Myr (left and right-hand panels) after the birth of a cluster containing a HMXB phase, with an ionization state corresponding to rows 1 and 2 of the middle column of Figure \ref{fig:NFW_density_ion_temp}. X-ray energies (blue dotted lines) are represented at 200eV, and UV energies (solid red lines) correspond to the threshold ionization energy for the relevant species: 13.6eV for hydrogen, 24.5eV for neutral helium, and 54.4eV for He$^+$. Additionally, $\tau = 1$ is plotted as a green dashed line for comparison. Please note that the initial rise of $\tau$ from zero is not visible in these plots because of size of the galaxy ($\sim 6$kpc) is too small to feature.

At 5 Myr, UV radiation is optically thin to hydrogen in the local IGM, but $\tau_{\rm H0}$ steadily increases with distance, surpassing 1 at $\sim$ 0.45 Mpc, and sharply rising to $\sim 10^3$ outside the initial ionization front. X-ray radiation, however, is optically thin to hydrogen throughout the originally ionized volume, and $\tau_{\rm H0}$ is only greater than 1 beyond 0.8 Mpc from the cluster. Even at early times, therefore, X-rays are able to propagate more freely through the ionized region, countering recombinations. After 50 Myr, both UV and X-ray radiation is optically thick to hydrogen. However, a small proportion of X-ray radiation ($\exp^{-\tau} \sim 10^{-7}$) is able to propagate through the IGM. The optical depth of UV radiation is a factor of $10^3$ higher, so it is almost completely confined within the host galaxy.  Although low, this fractional propagation of X-rays helps slow the recombination timescale for the ionized region. 

Similar results can be seen for helium, although $\tau_{\rm He0} \gg \tau_{\rm H0}$. Both UV and X-ray radiation are optically thick to He$^+$ at 5 and 50 Myr. However the X-ray optical depth is a factor of $\sim 50$ lower than the UV value, so a small fraction of He$^+$ can be ionized when HMXBs are included in the starburst model. 

Although X-rays enhance the lifetime of the ionized regions created by an initial burst of UV radiation, they do not extend the ionization front itself.

\subsubsection{Temperature Profiles}
The temperature plots in the right hand panel of Figure \ref{fig:NFW_density_ion_temp} show identical profiles at 5 Myr, exceeding $10^4$K in the innermost regions, and dropping off sharply at the edge of the ionization front. In the stellar-only model, as the ionized fraction rapidly falls, recombinations are the dominant cooling mechanism. However, the timescales for cooling are too long for thermal equilibrium to be reached. Therefore, although fully neutral within 50 Myr, the gas temperature remains at $\sim$$10^4$K, for the simulation lifetime. The partially ionized IGM that remains when HMXBs are included has a considerably larger electron fraction. Consequently, Inverse Compton cooling becomes the dominant cooling mechanism, followed closely by Bremsstrahlung cooling. Therefore, although the heating rates are increased when HMXBs are included, the gas temperature decreases at a greater rate than in the stellar-only case, dropping to $5{\times}10^3$ K after 100Myr\footnote{Note that the computed temperature is actually the electron temperature rather than the gas temperature.  The temperature of neutral hydrogen, for example, could be much lower if there has been insufficient time for it to reach equilibrium via scattering \citep[e.g.][]{1997ApJ...475..429M}.}.

\subsubsection{Comparison Between Spectra}

\begin{figure}
\centering
\includegraphics[width=0.5\textwidth]{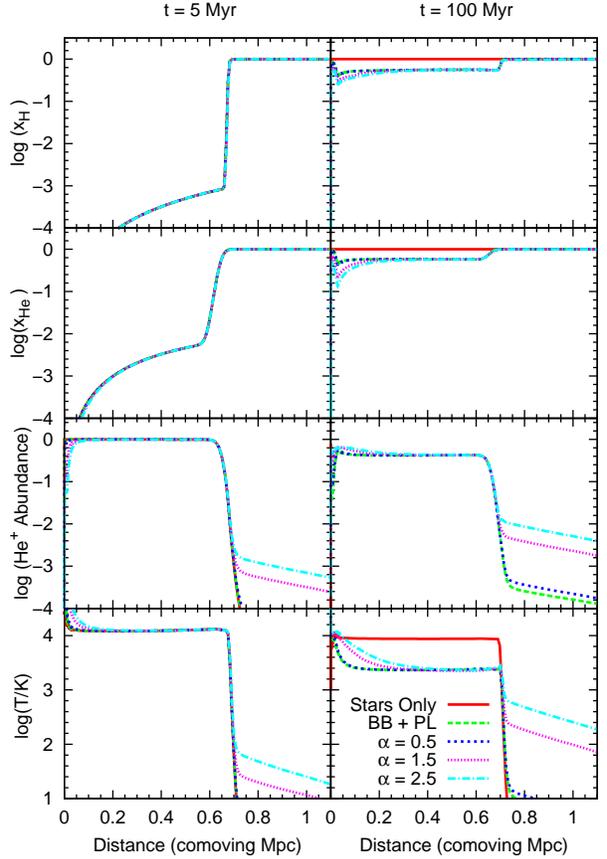}
\caption{\textbf{Heating and Ionization from a model stellar cluster embedded in a $\mathbf{10^8}$ M$_{\boldsymbol{\odot}}$ NFW halo: choice of HMXB spectrum.} Ionization state and temperature profiles of the gas surrounding five model stellar clusters containing (i) just stars, (solid red lines), (ii) both stars and HMXBs, assuming our fiducial blackbody and hard power law (BB+PL) HMXB spectrum, (green dashed lines) and, (iii-v) both stars and HMXBs, assuming a power law HMXB spectrum, with $\alpha = 0.5,1.5$ and $2.5$ (dark blue dotted, fine purple dotted, and light blue dot-dashed lines respectively). Profiles were calculated using a one dimensional radiative transfer code whereby radiation was propagated into a medium following an NFW density profile with M$_{\rm halo} = 10^8$ M$_{\odot}$. From top to bottom panels, hydrogen neutral fraction, helium neutral fraction, helium$^+$ abundance and gas temperature are plotted as a function of distance from the cluster. Panels in the left-hand column contain outputs from 5 Myr after the birth of the cluster, while those in the right-hand column plot the same for 100 Myr. }
\label{fig:NFW_density_compare_spectra}
\end{figure}

In rows 1-3 of Figure \ref{fig:NFW_density_compare_spectra}, the ionization profiles of hydrogen, helium  and He$^+$ are compared for four different HMXB spectra, at 5 Myr (left) and 100 Myr (right) after the birth of the stellar population, where the same NFW profile has been assumed. The stellar-only model is plotted as a solid red line and the BB+PL spectrum HMXB model is plotted as a green dashed line. In addition, clusters containing HMXBs with power law spectra, where $\alpha = 2.5,1.5$ and $0.5$,  are plotted as dark blue dashed, purple dotted and light blue dot-dashed lines respectively. At 5 Myr there is no difference in the ionization profiles of hydrogen or helium, as the ionizing output is dominated by main sequence stellar sources. After 100 Myr, however, some differences can be seen between HMXB spectra within 0.2 Mpc from the source. Models with soft spectra marginally increase the ionized fraction of material close to the source of radiation. The different shapes of the He$^+$ profiles are comparable to those in the constant density case (cf. Figure \ref{fig:const_density_compare_spectra}), with harder spectra ionizing a larger proportion of He$^{2+}$ ions. 

A comparison between the temperature profiles surrounding the  $10^{8}$ M$_{\odot}$ halo with different HMXB spectrum choices is shown in the bottom panels of Figure \ref{fig:NFW_density_compare_spectra}. Altering the HMXB spectra leads to broadly similar effects to those seen in the constant density case (Figure \ref{fig:const_density_compare_spectra}), with softer spectra producing higher local temperature peaks and increased heating beyond the main ionization front at late times.   

\section{Discussion}

Although we have shown that the inclusion of a HMXB phase in the evolution of a stellar cluster can extend the lifetime of its surrounding ionized volume, this effect is only visible when column densities are high enough to prevent the escape of UV radiation from the cluster. When all of the cluster's radiation is released directly into the IGM, X-rays from HMXBs have little effect on the ionization and heating of their surroundings. In both density profile choices, HMXBs were unable to extend the size of the ionization front. 

These results do not support the recent arguments in favour of HMXBs having a significant effect on their surroundings at high redshifts. X-rays from HMXBs do not ``pre-ionize'' distant regions or markedly raise temperatures, except in the case of particularly soft power-law HMXB spectra. In this section, we aim to explain these findings by answering the following two questions: firstly, why, in these scenarios, are we seeing such a negligible effect from HMXBs? Secondly, under what conditions could HMXBs have an important influence on their surroundings and are these constraints physically reasonable? 

\subsection{Why do HMXBs in our model have insignificant effects on their surroundings?}

\begin{figure*}
\centering
\includegraphics[angle = -90, width= 1.03\textwidth]{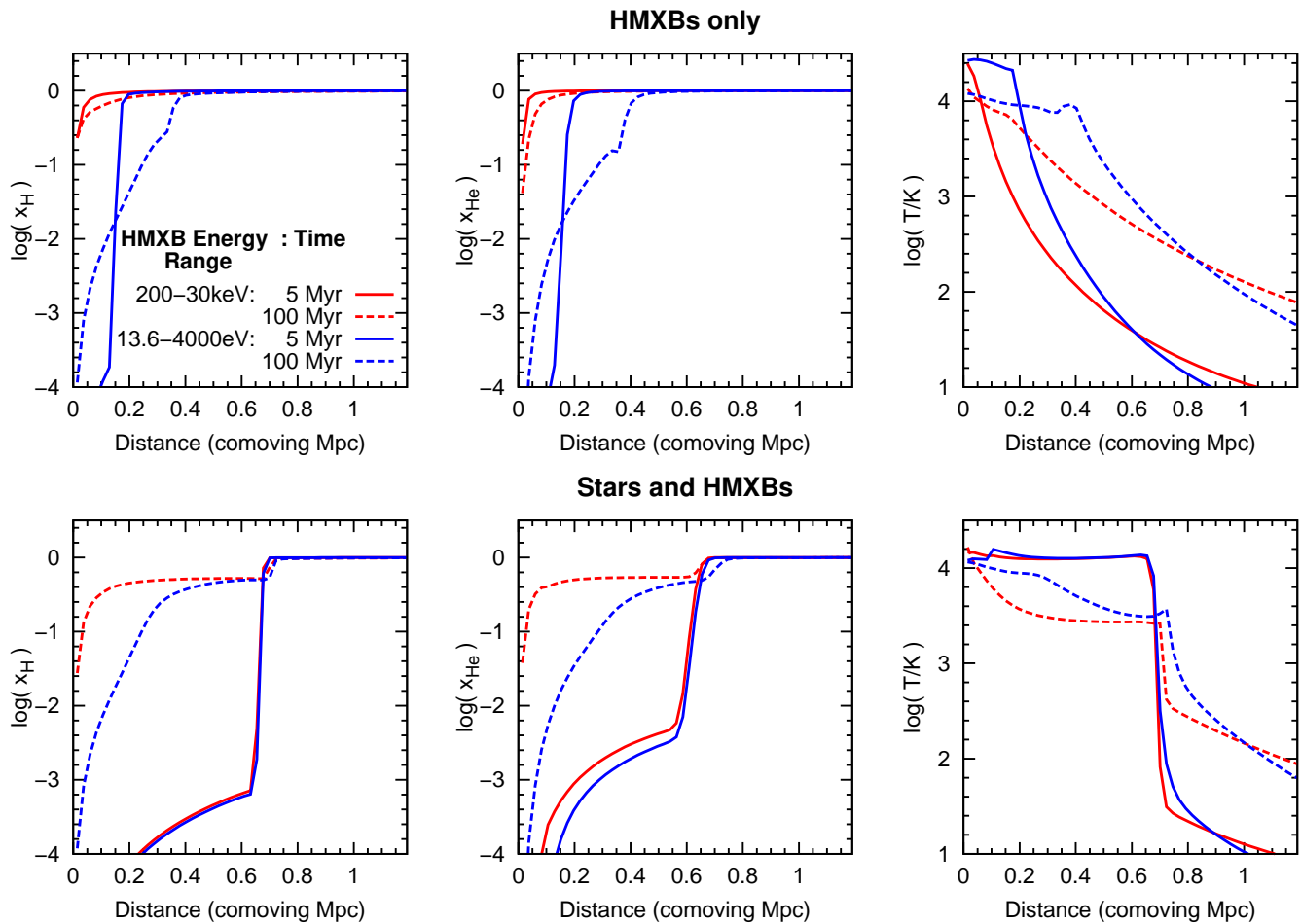}
\caption{\textbf{The effect of HMXB Spectral Energy Range on their Heating and Ionizing Potential.} Upper panels: Hydrogen neutral fraction, helium neutral fraction and temperature are plotted from left to right as a function of distance from the cluster for two model stellar populations emitting only HMXB radiation (the stellar component has been removed). Both models assume a power law HMXB spectrum, with $\alpha = 1.5$. They have 
the same integrated luminosity, but this is distributed over two different 
frequency ranges;  200eV- 30keV (red lines), and 13.6eV - 4keV (blue lines).  Results at 5 Myr and 100 Myr after the birth of the cluster are plotted as solid and dashed lines respectively. \newline
Lower Panels: Ionization and temperature profiles from the same simulations, but with stellar radiation included. These calculations have all been made using a one dimensional time dependent radiative transfer code, assuming a uniform, redshift-dependent, density medium. The models were run for 200 Myr, ending at z = 10. }
\label{fig:const_density_energies}
\end{figure*}

We have modelled the HMXBs in our stellar population synthesis with careful comparisons to observed systems and population statistics. These include making reasoned choices for HMXB spectra, lifetimes and abundances. These parameters all influence the predicted contribution of HMXBs to the radiation emitted by the cluster. The lack of additional ionization and heating of the IGM by the inclusion of HMXBs can therefore be discussed in the context of these selections. The main reasons for the insignificant influence of HMXBs on their surroundings can be explained as follows. 

\subsubsection{HMXB radiation is dwarfed by that of its progenitors}

The red lines in the upper panels of Figure \ref{fig:const_density_energies} show the effect of HMXB radiation on its surroundings when all main sequence stellar radiation is removed from the simulation. Here, we use a power law HMXB spectrum, normalised between 0.2-30keV, with $\alpha = 1.5$, as plotted in Figure \ref{fig:spectra}. From left to right, the plots show the hydrogen neutral fraction, helium neutral fraction, and gas temperature as a function of comoving distance. A uniform density, with the same conditions as those in Section \ref{sec:const_dens} has been assumed. The red solid lines show results at 5 Myr, and the red dashed lines show the profiles after 100 Myr. From these plots, we see that HMXBs can locally ionize hydrogen up to  $\sim 60\%$ ($\sim 90\%$), to distances of 100(400) kpc at 5 (100) Myr after the birth the cluster. They heat their local IGM to temperatures greater than $10^4$ K. The gas then cools slowly in the inner regions over time, but the temperature front moves outwards, raising the IGM temperature to 100K at  $> 1$ Mpc after 100 Myr. 

These results, if shown on their own, would predict a high influence of HMXBs on their surroundings. However, all HMXBs have stellar progenitors, and form with other stars. Therefore, in the context of a stellar cluster, rather than as an isolated source, HMXBs have relatively low influence. This can be seen in the red lines on the bottom panel of Figure \ref{fig:const_density_energies}, where the radiation of both HMXBs and main sequence stars is now included. In this case, the influence of stellar radiation, in particular the short-lived massive stellar progenitors of HMXB systems, is so high that the HMXB effect looks minimal and unimportant in comparison. This comparison warns against considering HMXBs out of context, as they would have been born into a region already fully ionized by their stellar progenitors.  

\subsubsection{HMXBs do not radiate in useful frequency ranges}

\begin{figure}
\centering
\includegraphics[width= 0.5\textwidth]{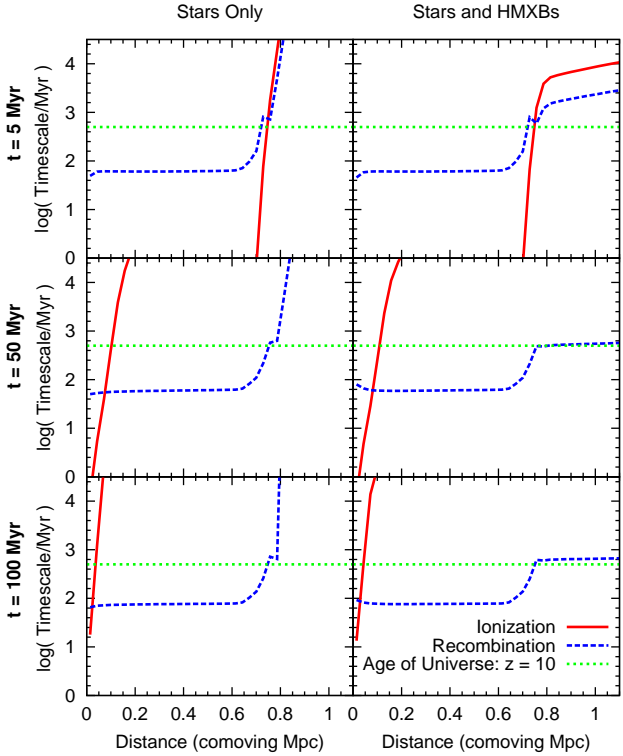}
\caption{\textbf{Ionization and Recombination Timescales.} Plotted in the upper, middle and lower panels  are the hydrogen ionization (solid red lines) and recombination (green dashed lines) timescales at 5, 50 and 100 Myr respectively after the birth of a model stellar cluster. Results are shown for a simulation containing stellar radiation only (left panels) and a model including both stars and HMXBs (right panels) assuming a BB+PL HMXB spectrum (see Figure \ref{fig:spectra}). These profiles were calculated using a one dimensional time dependant radiative transfer code over 200 Myr of the cluster's lifetime, assuming a uniform, redshift-dependent, density, with a final redshift of 10. Also plotted for comparison, as a dotted green line, is the age of the universe at z = 10.}
\label{fig:const_density_timescales}
\end{figure}

Also plotted in blue in Figure \ref{fig:const_density_energies} are the neutral hydrogen, neutral helium and temperature profiles of a constant density IGM, when the HMXB energy is emitted over an alternative frequency range. In the upper panels, the main sequence stellar radiation has been removed. While the integrated luminosity is the same as for the previous model, the HMXB energy is instead normalised between 13.6eV - 4keV, the range used by our radiative transfer code. In this case, a much larger proportion of HMXB energy is in the useful ionizing range.  As shown in the top left panel, HMXBs with these spectra could fully ionize hydrogen to a distance of 100 kpc after 5 Myr (solid blue line). The ionized zone then expands over time, with material out to 400 kpc remaining significantly ionized at 100 Myr (blue dashed line). Similarly, the helium ionization front extends to 100 kpc at 5 Myr, and propagates further as material behind the front slowly recombines. Local temperatures are also higher with this lower frequency spectrum.

In the blue lines on the lower figures, HMXB radiation is also moved into the 13.6eV - 4keV range, but the main sequence stellar component is included. Lowering the frequency of HMXB radiation results in higher hydrogen and helium ionized fractions, particularly after 100 Myr. The ionization fronts also extend marginally further. After 100 Myr, the temperatures are also higher, within $\sim 1$ Mpc for this altered HMXB spectrum. 

These plots highlight the importance of our spectrum choice. The fact that HMXBs emit energy in X-rays ($> 200$eV) is an important factor in determining their ionizing and heating potential. The long mean free paths of these photons mean they do not add much to the ionizing and heating influence of stellar cluster, despite their high luminosities. 

\begin{figure*}
\centering
\includegraphics[angle = -90,width= 1.05\textwidth]{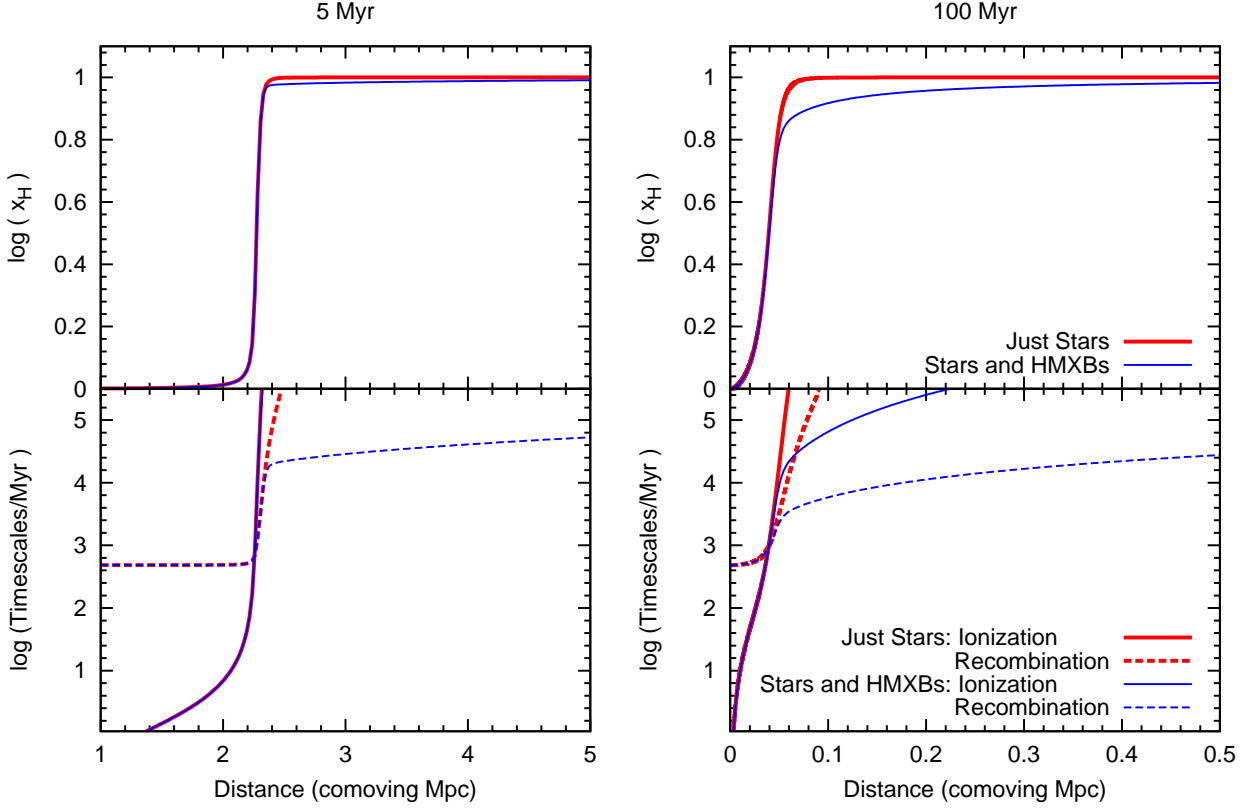}
\caption{\textbf{Ionization profiles and timescales determined using a static photoionization-recombination equilibrium calculation.} In the top left-hand panel, the hydrogen neutral fraction is plotted against comoving distance from two model stellar clusters, without (solid red line) and with (solid blue line) a HMXB phase. These profiles were calculated by solving a static photoionization-recombination equilibrium calculation after inputting the radiation emanating from the clusters at 5 Myr. The top right panel shows solutions to the same calculations at 100 Myr. On the bottom row, the static photoionization and recombination timescales are plotted for each model as a function of distance from the cluster at 5 Myr and 100 Myr on the left and right-hand panels respectively. The stellar-only model is plotted in red, and the model comprising both stars and HMXBs is plotted in blue, with solid lines for photoionization timescales, and dashed lines for recombination timescales.}
\label{fig:static_ion_timescales}
\end{figure*}

\subsubsection{Ionization Timescales are Too Long}

The reason that X-rays from HMXBs have little effect on their surroundings can be seen in Figure \ref{fig:const_density_timescales}.  Here, the hydrogen ionization and recombination timescales for the models in Figure \ref{fig:const_density_ion_temp} have been calculated as:
\begin{align}
 &\rm Photoionization \; \rm Timescale:\; t_{\rm ion} = \Gamma^{-1}\\
& \rm Recombination \; \rm Timescale:\; t_{\rm rec} = (\alpha n_e)^{-1}
\end{align}
Where $\Gamma$ is the photo-ionization rate (s$^{-1}$), $\alpha$ is the recombination rate (cm$^3$s$^{-1}$) and $n_e$ is the electron density ($cm^{-3}$) as defined in Appendix B4 of \citet{2007MNRAS.374..493B}.

The left column shows the stellar-only case, whereas a model containing a HMXB phase with a BB+PL spectrum is plotted in the right-hand panels. These columns correspond to the left and middle panels of Figure \ref{fig:const_density_ion_temp} respectively. Ionization timescales are plotted as solid red lines and recombination timescales are plotted as blue dashed lines. These are shown at 5, 50 and 100 Myr after the birth of the cluster in the top, middle and bottom panels respectively. Also plotted, for context, as a horizontal green dotted line, is the age of the universe at z=10 ($\approx$ 500 Myr), which was calculated for $H_0 = 0.7$, $\Omega_M = 0.27$ and $\Omega_{\rm vac} = 0.73$ using the web-based cosmology calculator described in \citet{2006PASP..118.1711W} 

At 5 Myr, for a main-sequence stellar radiation only model (left-hand panels), the ionization timescale is extremely short ($<$ 1 Myr) out to $\sim$ 0.7 Mpc -- the edge of the fully ionized volume -- before rising sharply to $>$ 10 Gyr. The recombination rate, which depends on the number of free electrons, is $\sim$ 60 Myr behind the ionization front and also increases rapidly at 0.7 Mpc. The timescale profiles for recombination do not change significantly in the next 100 Myr. However, 50 Myr into the cluster's lifetime, only the inner 100kpc has an ionization timescale of $<$ 500 Myr, and after 100 Myr, the front has receded to just 50 kpc. Ionization is therefore slower than recombination at almost all distances within the initially ionized region, so it slowly recombines. 

The right-hand profiles show the equivalent results for a cluster containing both stars and HMXBs. Within 0.7 Mpc, the ionization and recombination timescales at 5, 50 and 100 Mpc are very similar to the stellar-only case. However, at 5 Myr during the era of the most luminous HMXBs, both the ionization and recombination timescales are reduced, to $\sim$ 1 Gyr at the edge of the ionization front, gradually rising at greater distances. The ionization timescale is approximately two times longer than the recombination timescale. This implies that, after $>10^3$ Myr, an equilibrium between reionization and recombination could result in low level partial ionization of the distant IGM. However, this equilibrium cannot be reached in the lifetime of the cluster, so this solution could never form. 

For comparison, we solve a static version of the photoionization-recombination equations, to demonstrate the equilibrium solution that can be derived when timescales are not accounted for. For this purpose, we have used the calculation described in \citet{2005MNRAS.360L..64Z,2007MNRAS.375.1269Z}, which we write in the form:
\begin{equation}
\alpha_{\rm HI}^{(2)}n_{\rm H}^2(1-x_{\rm HI})^2 = \Gamma( r)n_{\rm H}x_{\rm HI} \label{eqn:static}
\end{equation}

\noindent Here, the case-B recombination coefficient, $\alpha_{\rm HI}^{(2)}$, is equal to $2.6\times 10^{-13}(T /10^4 K )^{-0.85} \rm cm^3 \rm s^{-1}$ \citep[cf. ][]{1970masp.book.....K}, for a gas
temperature $T$, where we use $T = 10^4K$. We evaluate the mean number density of hydrogen, $n_H\sim 1.9 {\times} 10^{-7}(1+z)^3\rm cm^{-3}$  \citep{2007ApJS..170..377S} at $z = 10$. The ionization rate per hydrogen atom at distance $r$ from the source is given by 

\begin{equation}
\Gamma( r) = \int_{E_0}^{\infty}\sigma(E)\mathcal{N}(E;r)\left[1+\frac{E}{E_0}\phi(E,x_e)\right]\frac{dE}{E}
\end{equation}

\noindent where  $\sigma_H(E) = \sigma_0(E/E_0)^{−3}$ is the bound-free absorption cross-section for hydrogen, with $\sigma_0 = 6\times 10^{-18}\rm cm ^2$ and $E_0 = 13.6 \rm eV$. $\mathcal{N}(E;r)$ is the number of photons per unit time per unit area at a distance $r$ from the source.

\begin{equation}
\mathcal{N}(E,r) = \rm e^{{-\tau(E;r)}}\frac{F(E)}{4{\pi}r^2} \rm cm ^{-2} \rm s ^{-1}
\end{equation}

\noindent where the optical depth is evaluated as

\begin{equation}
\tau(E;r) = \int_0^r n_{\rm H}x_{\rm HI} \sigma(E) dr
\end{equation}

\noindent and the spectrum, F(E), is inputted from our Monte Carlo model of the stellar
population. The term  $[1+(E/E_0)\phi(E,x_e)]$ accounts for secondary ionizations, where

\begin{equation}
\phi(E,x_e) =0.3908[1-(1-x_e^{0.4092})^{1.7592}]
\end{equation}

\noindent and $x_e = 1- x_H$ \citep{1985ApJ...298..268S}.

These equations are solved at 5 and 100 Myr, in the left and right-hand panels of Figure \ref{fig:static_ion_timescales}. Note that the fully ionized zones are overestimated because the spectrum at $r$ is calculated as  $F(E)\exp^{- \tau}/4{\pi}d^2$, so that the photons used to ionize the material within $r$ have not been removed. Furthermore, ionization history has not been accounted for, in the equilibrium calculation, which is why the ionized volume appears to recede over time, rather than recombining as a whole.

The upper panels of Figure \ref{fig:static_ion_timescales} show the hydrogen neutral fraction against time, while the lower panels show the ionization and recombination timescales. In the upper panels, the stellar-only model is plotted in red, and a model with a HMXB phase, using a BB+PL HMXB spectrum, in shown in blue. At both 5 and 100 Myr, the addition of HMXB radiation leads to low level ionization of the distant IGM, beyond the fully ionized volume produced by the main sequence stellar radiation. 

\begin{figure}
\centering
\includegraphics[width= 0.5\textwidth]{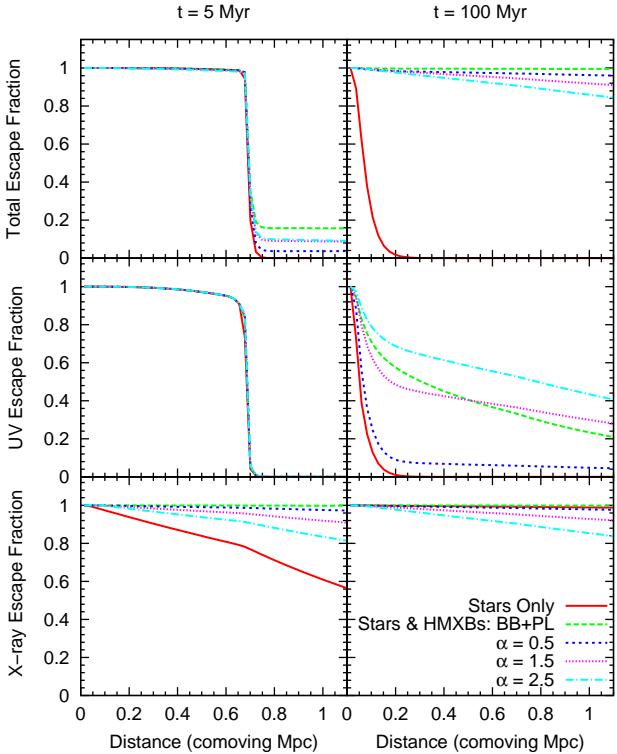}
\caption{\textbf{Heating and Ionization from a model stellar cluster surrounded by a uniform density medium: escape fractions.} The escape fraction; the fraction of energy in a given frequency range that remains unabsorbed, is plotted as a function of distance from the cluster for different stellar population models. A constant redshift-dependent density profile has been assumed, and the calculations have been made using our one dimensional radiative transfer code, over a cluster lifetime of 200 Myr, ending at z = 10. The upper panels show the total 13.6eV-4keV escape fraction for five different models at 5 Myr (left panel) and 100 Myr (right panel) after the birth of the cluster. The models plotted are (i) stellar radiation only (solid red line), (ii) stars and HMXBs, with a blackbody + hard power law HMXB spectrum (our fiducial model; green dashed line) and (iii-iv) power law models with $\alpha = 0.5,1.5$ and $2.5$ (dark blue dotted, fine purple dotted and light blue dot-dashed lines respectively). The panels on the second and third rows split this escape fraction into UV (13.6-200eV) and X-rays (200-4000eV) respectively. }
\label{fig:const_density_escape_fractions}
\end{figure}

The lower panels show the ionization (solid lines) and recombination (dashed lines) timescales for the two different models. The red lines correspond to the stellar-only model, and the blue to the model comprising both stars and HMXBs. In the stellar-only case, both the photoionization and recombination timescales rise rapidly beyond the ionization front, and negligible partial ionization is visible in the upper panels. However, at late times, when HMXBs are included, both the recombination and photoionization timescales are reduced beyond the main ionized volume. Here, at 100 Myr and a distance of 0.1Mpc,  the photoionization rate is only a factor of 10 higher than the recombination rate. This accounts for the partially ionized equilibrium visible in the upper panels. 

Of course, all the timescales outside the ionization fronts are longer than $\sim$ 500 Myr, the age of the universe at z = 10, which implies that the partially ionized equilibrium at these distances is unreachable. We include this discussion to highlight the limitations of simple calculations and general arguments used to argue in favour of alternative or additional ionizing sources. A boost in ionizing energy does not necessarily translate to increased ionization, as the spectrum of the new source needs to be considered; in the context of our isolated cluster model, X-rays may require too long a timescale to have a significant influence because the sources are short lived.

\subsection{Can HMXBs have an Important Feedback Effect?}

We have shown that, in the context of our starburst model, radiation from HMXBs does not produce significant additional ionization and heating of the IGM, either in a constant or varying density medium. Indeed, in the NFW density case, including HMXBs can only prolong the lifetime of previously ionized structures, which actually results in lower IGM temperatures. We now discuss the fate of the HMXB photons in our models, and the conditions required for their feedback effects to be non-negligible.

\subsubsection{Escape Fractions}

\begin{figure*}
\centering
\includegraphics[angle = -90, width= 1.1\textwidth]{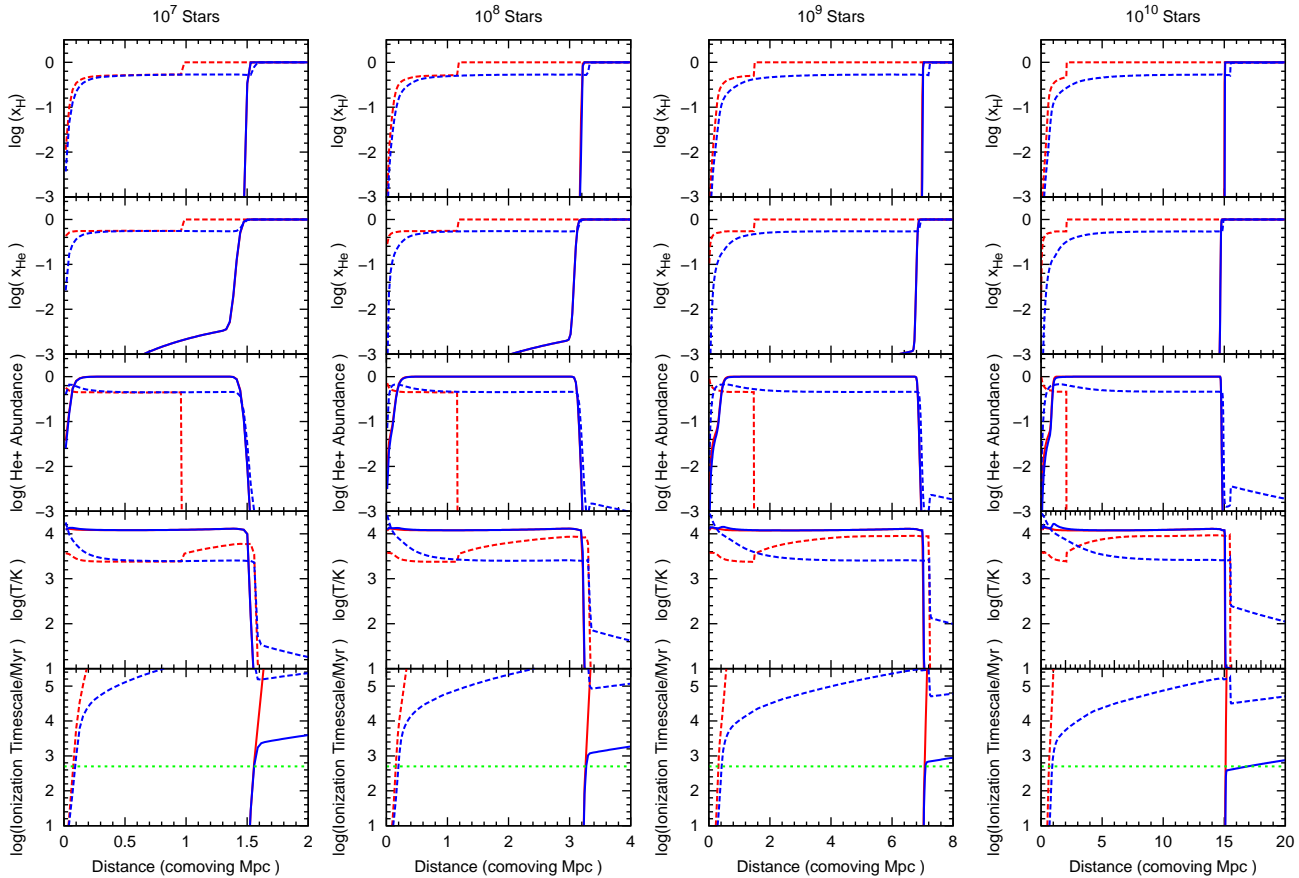}
\caption{\textbf{Heating and Ionization from multiple clusters in a uniform density medium: boosting HMXB ionization and heating power.} From left to right, we use our  one dimensional radiative transfer code to calculate the ionization and temperature profiles surrounding 10, 100, 1000, and 10000 clusters. Their combined emission is propagated through a constant, redshift-dependent, density medium, for 200 Myr of the cluster lifetime, ending at z = 10.  From top to bottom, the hydrogen neutral fraction, helium neutral fraction, helium$^+$ abundance and gas temperatures have been plotted as a function of distance from the clusters for two models; without (red lines) and with (blue lines) HMXBs. Solid lines show the profiles at 5 Myr after the birth of the cluster, and dashed lines show the profiles after 100 Myr. Note that the horizontal scales are different in each column. }
\label{fig:const_density_boost}
\end{figure*}

One of the arguments in favour of an X-ray contribution to reionization cites their long mean free paths, compared to UV photons \citep{2009MNRAS.395.1146P,2011A&A...528A.149M}. This means photons are able to escape further into the IGM rather than being absorbed in their immediate surroundings. 

In the upper panels of  Figure \ref{fig:const_density_escape_fractions}, we plot the escape fractions as a function of distance for our constant density IGM model, where the escape fraction is defined as

\begin{equation}
f_{\rm esc}(r ) = \frac{\int_{13.6\rm eV}^{4\rm keV}F(E;r)dE}{\int_{13.6\rm eV}^{4\rm keV}F(E;0)dE}\rm{,}
\end{equation}

i.e. the proportion of ionizing energy emitted from the cluster that reaches a distance $r$. 

Five models are plotted; a stellar-only simulation (solid red line) and four models including a HMXB phase with differing HMXB spectra; a BB+PL spectrum (dashed green line) and power law spectra with $\alpha = 2.5$ (dark blue dashed line), 1.5 (pink dotted line) and 0.5 (light blue dot-dashed line). The escape fractions are plotted at 5 Myr and 100 Myr after the birth of a stellar cluster, in the left and right-hand panels. On the second and third rows, this escape fraction is split into a UV (13.6eV-200eV) and X-ray  (200eV-4keV) component respectively. 

At 5 Myr, the total escape fractions drop to below 20$\%$ beyond the ionization front for all models, as radiation from main sequence stars dominates the cluster SED. In the middle panel, it is clear that all of the UV energy is absorbed within this region. However, the bottom panel shows that 90$\%$ of the X-ray energy escapes for all models including HMXBs, with more energy escaping when the HMXB spectrum is harder. Note that the X-ray radiation from main sequence stars is negligible, so its divergence from the other lines in the bottom panel is unimportant.

After 100 Myr, radiation from HMXBs dominates the ionizing emission from the cluster, so more pronounced differences are seen in the total escape fractions between models with and without a HMXB phase.  In the stellar-only model, the escape fraction decreases sharply with distance from the source, with all energy absorbed within 0.2 Mpc. In the models involving a HMXB phase, however, most of the energy escapes, because the SED is much harder. The UV escape fraction is greater for harder spectra, and the X-ray escape fraction remains similar to the 5 Myr snapshot, with $>80\%$ of the X-rays escaping beyond 1 Mpc in all models including a HMXB phase.

These escape fraction plots show that, as theory suggests, a significant proportion of HMXB energy penetrates deep into the IGM. It does not make an impression on the ionization state of the material at these distances because of long ionization timescales. However, should the X-ray background build up to greater levels, these timescales could be reduced and an effect may be visible.

\subsubsection{Greater HMXB Number Densities}

We have established that the ionization timescales for the X-rays from HMXBs are too long for them to exert a major effect on the ionization state of their distant surroundings, despite their high escape fractions. A higher number density of HMXBs might lower these timescales. This could be achieved in three ways; 

\begin{enumerate}
\item increasing the HMXB to main sequence star ratio,
\item modelling a larger starburst, so that more HMXBs can form,
\item continuous star formation, allowing for more uniform X-ray production (see Section 4.4).
\end{enumerate}

Our estimates of the proportion of stellar binaries that survive the primary's supernova phase to become HMXBs ($30\%$) are observationally motivated. Additionally, we have assumed $f_{\rm sur} = 1$ throughout this paper, thus neglecting any other evolutionary reasons for HMXBs failing to form. Therefore, increasing the ratio of HMXBs to main sequence stars, as in option (i), would be implausible within the context of our simulations.

In Figure \ref{fig:const_density_boost} we investigate (ii) by increasing the size of our stellar clusters from $10^6$ to $10^{10}$ stars. Here, the total energy from the cluster with and without a HMXB phase is propagated through a constant density medium, as described in Section \ref{sec:const_dens}. From top to bottom, the hydrogen neutral fraction, helium neutral fraction, He$^+$ abundance, temperature and ionization timescale are plotted against distance for a stellar-only model (red lines) and a stellar and HMXB model, with a BB+PL HMXB spectrum and $f_{\rm sur }= 1$ (blue lines). These are plotted at 5 Myr after the birth of the cluster (solid lines), and 100 Myr (dashed lines).  Four different cluster sizes are shown from left to right, containing $10^7$, $10^8, 10^9$ and $10^{10}$ stars respectively. Note that the horizontal scale is increased for more massive clusters, to incorporate the larger ionized volumes.
 
Since a greater X-ray background density results in shorter ionization timescales, the relative influence of stars and HMXBs on their surroundings is non-linear. Therefore, although stars dominate the heating and ionizing influence of the cluster on the IGM, HMXBs show more significant effects as the cluster size increases.  In a $10^7$ star cluster, the hydrogen ionization profiles (top left plot) show little difference at 5 Myr. However, although the ionization front is not extended at this time, after 100 Myr a partially ionized zone remains that is marginally larger than the originally ionized region. There is no matching increase in the volume of ionized helium, but recombinations are also quenched by HMXB radiation. As the number of stars in the cluster increases, towards the right-hand panel of Figure 10, the effect of HMXBs on their environment increases relative to that of the main sequence stellar component.  In particular, the extent of the partially ionized regions after 100 Myr increases. However, at no point is the outer edge of the ionization front extended because the ionization timescales (bottom panels), while much reduced by the inclusion of HMXBs, remain prohibitively long. 
 
The temperature at 5 Myr is dominated by stellar radiation for all cluster sizes, but the effects of HMXBs are increasingly significant as cluster size increases. Close to the cluster, and beyond the ionization front, a HMXB phase results in higher temperatures. However, within the main ionized region the higher cooling rate of the partially ionized regions implies that including HMXBs results in lower temperatures, as seen in the NFW density results of Figure \ref{fig:NFW_density_ion_temp}. 

\subsection{Mechanical Feedback and Local Effects}

We note that HMXBs may be able to exert influence on the IGM via alternative mechanisms that have not been considered in this study. For example, several ULXs are known to drive large scale outflows \citep[e.g. SS 433;][]{2001ApJ...562L..79B}. These alternative feedback processes are addressed in \citet{2012MNRAS.423.1641J}. Furthermore, while this study may rule out any strong feedback effect of HMXBs on the IGM, we have not addressed the radiative and mechanical effects that HMXBs may exert on their host galaxies. Significant feedback within these dense environments may be plausible. 

\subsection{Further Potential Influence on the IGM}

In our model, we treat star formation as a single burst over a short timescale. While this may be relevant for e.g. globular clusters, star formation is likely to be a  continuous  process within young galaxies. HMXBs are unable to sustain a sufficient ionizing background for significant ionization and heating under these conditions, due to their brief lifetimes. However, ongoing HMXB formation may be able to maintain a low level X-ray ionizing background, pre-ionizing and heating the distant IGM before UV radiation from main sequence stellar sources penetrates these distances.

A further consideration is the fate of hard X-ray photons, with energies greater than the 4keV limit used in our models. A neutral IGM is optically thin over a Hubble length to photons with energies greater than $E_{\rm lim} =  1.8[(1+z)/15]^{0.5}$keV, where $E_{\lim} \simeq 1$keV at $z=10$ \citep{2004ApJ...613..646D}.  Hence, it is extremely unlikely that hard X-ray photons would have been absorbed in their immediate surroundings.  Consequently, these photons are unimportant for the photoionisation calculations, within a few comoving Mpc of HMXBs, that we present in this work.

Note, however, that recent (model dependent) constraints on the 21cm brightness temperature by PAPER-32 imply that the IGM may have been partially heated by X-ray sources at redshift $z=7.7$ (Parsons et al. 2014).  Fast electrons produced by hard X-ray and gamma ray photons could lead to cascades, resulting in multiple secondary ionizations.  For a 1 MeV photon, 32\% of the energy is transferred to the IGM via ionization \citep{2010MNRAS.404.1569V}.   As a significant proportion of the radiation from HMXBs falls above $E_{\rm lim}$  \citep[e.g.][]{1999MNRAS.309..496G}, it remains possible that subsequent high energy cascades may heat the neutral, pre-reionisation IGM to a low level on large scales.

\subsection{Conclusions}

We have investigated the influence of HMXBs on the high redshift IGM, using our model coeval stellar population, comprising main sequence stars formed in a single burst followed by a later HMXB phase. With observationally motivated choices for HMXB abundances, spectra and lifetimes, we argue that our models provide a good estimate of HMXB emission in a plausible context. We used a one dimensional radiative transfer code to predict the ionization and temperature profiles surrounding the population. We have considered the interaction of the stellar radiation from our model cluster through a uniform density IGM from z = 14.5 to z = 10, and through a $10^8 M_\odot$ one-dimensional NFW profile, with and without a HMXB phase. 

For a constant density IGM, HMXBs produce negligible enhanced ionization, except in the case of extreme HMXB spectra. This is because their high energy SEDs and brief lifetimes leave insufficient time for an X-ray ionized background to build up. Radiation from massive main sequence stars raises the temperature to $10^4$ K, but including a HMXB phase only increases the heating rate in the inner $\sim  0.1$ Mpc. In the NFW profile case, UV radiation from the most massive stars initially ionizes a large IGM volume, which later recombines as the reduced stellar radiation is unable to penetrate the high density galactic core. HMXB photons stall recombinations behind the front, keeping it partially ionized for longer, but do not extend the outer edge of the ionized zone. The partially ionized regions have an increased number of free electrons, so enhance both Inverse Compton and Bremsstrahlung cooling. Consequently, HMXBs actually increase the cooling rate in these regions, halving the gas temperature after 100 Myr. In neither case is thermal equilibrium reached during the simulation’s lifetime, so cooling is ongoing.

Although we have shown that HMXBs do not have a strong influence on the IGM in the context of our starburst model, they may nevertheless have a significant local influence within their host galaxies. It could be the case that HMXBs can influence the IGM through factors that are beyond the scope of our models: e.g. it is possible that multiple ionizations from hard X-ray photons, via high energy cascades, may have deposited  heat and ionizing energy into the environment in addition to that seen in the models as presented here. However, we view this possibility as unlikely given the very low optical depths in the IGM for photons of energy $\gsim$  1 keV. A more significant caveat is likely to be the assumption of HXMB formation via a starburst. Ongoing or persistent HMXB formation may be able to maintain a low level X-ray ionizing background, which could partially ionize the distant IGM. 

\section*{Acknowledgements}

GK acknowledges the support of an STFC studentship. CP acknowledges the support of Australian Research Council Future Fellowship FT130100041. JSB acknowledges the support of a Royal Society University Research Fellowship. This research is supported by an STFC consolidated grant. 

\bibliographystyle{mn2e}
\bibliography{HMXBs_early_universe.bib}

\label{lastpage}

\end{document}